\documentclass[twocolumn,showpacs,preprintnumbers,amsmath,amssymb]{revtex4}

\usepackage{graphicx}

\begin{document}
\title{Quantum critical behavior of ultracold atoms in two-dimensional optical lattices}

\author{Xibo Zhang}
\author{Chen-Lung Hung}
\author{Shih-Kuang Tung}
\author{Cheng Chin}
\address{The James Franck Institute and Department of Physics, \\ The University of Chicago, Chicago, IL 60637, USA}
\date{\today}

\begin{abstract}

As the temperature of a many-body system approaches absolute zero, thermal fluctuations of observables cease and quantum fluctuations dominate. Competition between different energies, such as kinetic energy, interactions or thermodynamic potentials, can induce a quantum phase transition between distinct ground states. Near a continuous quantum phase transition, the many-body system is quantum critical, exhibiting scale invariant and universal collective behavior~\cite{Coleman05Nat, Sachdev99QPT}. Quantum criticality has been actively pursued in the study of a broad range of novel materials~\cite{vdMarel03Nat, Lohneysen07rmp, G08NatPhys, Sachdev08NatPhys}, and can invoke new insights beyond the Landau-Ginzburg-Wilson paradigm of critical phenomena~\cite{Senthil04prb}. It remains a challenging task, however, to directly and quantitatively verify predictions of quantum criticality in a clean and controlled system. Here we report the observation of quantum critical behavior in a two-dimensional Bose gas in optical lattices near the vacuum-to-superfluid quantum phase transition. Based on \textit{in situ} density measurements, we observe universal scaling of the equation of state at sufficiently low temperatures, locate the quantum critical point, and determine the critical exponents. The universal scaling laws also allow determination of thermodynamic observables. In particular, we observe a finite entropy per particle in the critical regime, which only weakly depends on the atomic interaction. Our experiment provides a prototypical method to study quantum criticality with ultracold atoms, and prepares the essential tools for further study on quantum critical dynamics.

\end{abstract}






\maketitle

In the vicinity of a continuous quantum phase transition, quantum fluctuations lead to non-classical universal behavior of a many-body system~\cite{Coleman05Nat}. Quantum criticality not only provides novel routes to new material design and discovery~\cite{Coleman05Nat}, but also raises possible links between condensed matter systems and those studied in nuclear physics~\cite{Stoof01Nat, Senthil04sci} or in cosmology~\cite{Coleman05Nat,Sachdev09JPCM}. Understanding quantum criticality and its general role in strongly correlated systems has hence attracted significant studies such as those on heavy-fermion materials~\cite{G08NatPhys, L94prl}, quasi-one-dimensional Ising ferromagnets~\cite{Coldea10sci}, quantum antiferromagnets~\cite{Boehm08prl}, ruthenate metals~\cite{Grigera01sci}, and chromium at high pressure~\cite{Rosenbaum10PNAS}.

Ultracold atoms offer a clean system for a quantitative and precise study of quantum phase transitions and critical phenomena~\cite{Greiner02Nat, Tilman10Nat,HCN10Nat, Xibo11njp, Simon11Nat}. As an example, the superfluid-to-Mott insulator quantum phase transition is realized by loading atomic Bose-Einstein condenstates into optical lattices~\cite{Greiner02Nat}. In recent experiments, universal scaling behavior was observed in interacting Bose gases in three \cite{Esslinger07sci} and two dimensions~\cite{CL11Nat}, and in Rydberg gases~\cite{Low09PRA}. Suppression of the superfluid critical temperature near the Mott transition was observed in 3D optical lattices~\cite{Trotzky10NatPhys}. Studying quantum criticality in cold atoms based on finite-temperature thermodynamic measurements, however, remains challenging and has attracted increasing theoretical interest in recent years~\cite{Qi10prl,Kaden10,DW10pra,Nandini08NatPhys}.

In this letter, we report the observation of quantum critical behavior of ultracold cesium atoms in a two-dimensional (2D) optical lattice across the vacuum-to-superfluid transition. This phase transition can be viewed as a transition between Mott insulator with zero occupation number and superfluid, and can be quantitatively described by the Bose-Hubbard model~\cite{Fisher89prb}. Our measurement is based on atomic samples near a normal-to-superfluid transition, which connects to the vacuum-to-superfluid quantum phase transition in the zero-temperature limit. At progressively lower temperatures, quantum criticality is revealed in the emergence of universal scaling of the equation of state. From the equation of state, we extract the quantum critical point and the critical exponents, and compare them with theoretical predictions. Furthermore, we observe the breakdown of quantum criticality at high temperatures and estimate an upper thermal energy scale for the quantum critical behavior. The derived scaling laws permit a complete determination of the thermodynamics of the critical gas. In particular, we observe a universal, non-zero entropy per particle in the critical regime, carrying a weak dependence on the atomic interactions.

\begin{figure}[t]
\includegraphics[width=3.4 in]{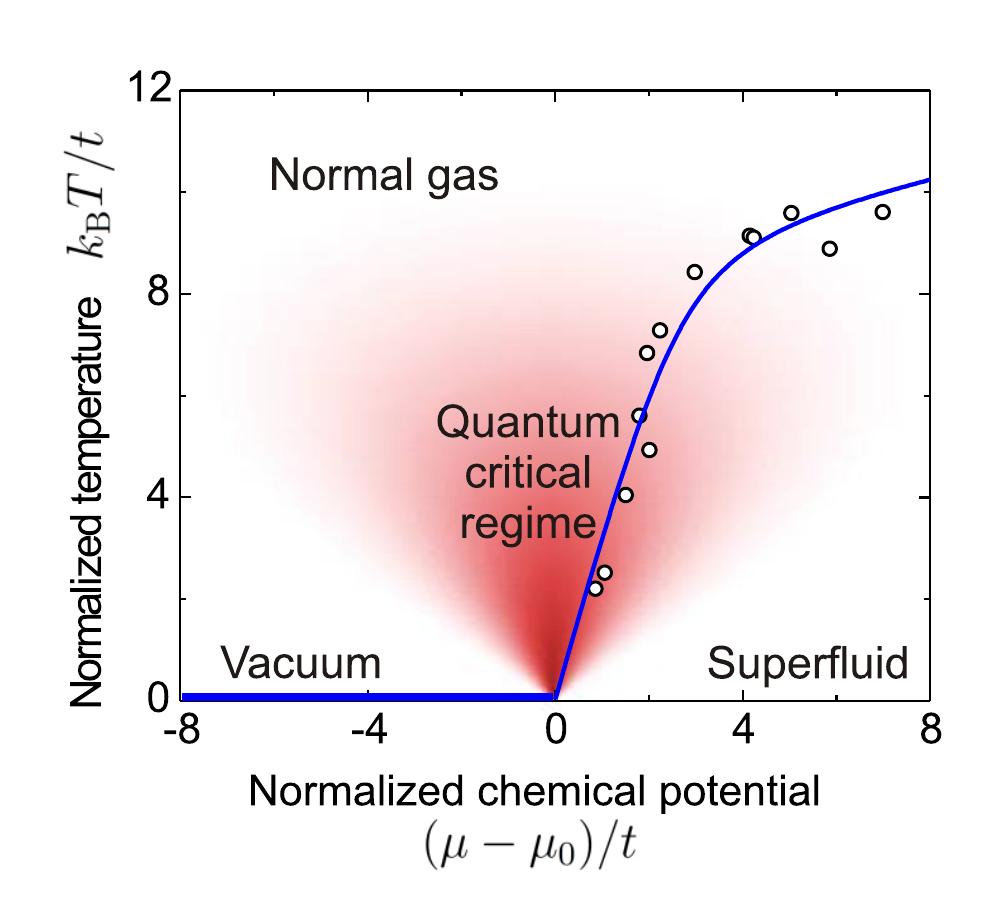}
\caption{\textbf{The vacuum-to-superfluid quantum phase transition in 2D optical lattices.} At zero temperature, a quantum phase transition from vacuum (horizontal thick blue line) to superfluid occurs when the chemical potential $\mu$ reaches the critical value $\mu_{0}$. Sufficiently close to the transition point $\mu_{0}$, quantum criticality prevails (red shaded area), and the normal-to-superfluid transition temperature $T_{\mathrm{c}}$ (measurements shown as empty circles; see Methods) is expected to vanish as $T_{\mathrm{c}} \sim (\mu-\mu_{0})^{z\nu}$; the blue line is a guide to the eye. From the prediction $z\nu = 1$~\cite{Fisher89prb,Kaden10,DW10pra}, the linearly extrapolated critical chemical potential is $\mu_{0} = -3.6(6)t$, consistent with the theoretical value $-4t$~\cite{Xibo11njp}. Here both the thermal energy scale $k_{\mathrm{B}}T$ and chemical potential $\mu$ are normalized by the tunneling $t$.}\label{fig1}
\end{figure}

The quantum phase transition and quantum critical regime in this study are illustrated in Fig.~\ref{fig1}. The zero-temperature vacuum-to-superfluid transition occurs when the chemical potential $\mu$ approaches the quantum critical point $\mu_{0}$. Sufficiently close to the quantum critical point, the critical temperature $T_{\mathrm{c}}$ for the normal-to-superfluid transition is expected to decrease according to the following scaling~\cite{Fisher89prb}:
\begin{equation}
\frac{k_{\mathrm{B}}T_{\mathrm{c}}}{t} = c\left(\frac{\mu - \mu_{0}}{t}\right)^{z\nu},
\end{equation}
where $k_{\mathrm{B}}$ is the Boltzmann constant, $t$ is the tunneling energy, $z$ is the dynamical critical exponent, $\nu$ is the correlation length exponent, and $c$ is a constant. In the quantum critical regime~(shaded area in Fig.~\ref{fig1}), the temperature $T$ provides the sole energy scale, and all thermodynamic observables are expected to scale universally with $T$~\cite{Fisher89prb}. Thus the equation of state is predicted to obey the following scaling~\cite{Qi10prl,Kaden10,DW10pra}
\begin{equation}\label{EoS}
\tilde{N}  =  F(\tilde{\mu}),
\end{equation}
in which $F(x)$ is a universal function, and 
\begin{equation}\label{DefScaled}
\tilde{N} = \frac{N - N_{\mathrm{r}}}{(\frac{k_{\mathrm{B}}T}{t})^{\frac{D}{z}+1-\frac{1}{z\nu}}} \textrm{  and  }  \tilde{\mu} = \frac{\frac{\mu-\mu_{0}}{t}}{(\frac{k_{\mathrm{B}}T}{t})^{\frac{1}{z\nu}}}
\end{equation}
\noindent are the scaled occupation number and scaled chemical potential, respectively. Here $N$ is the occupation number, $D$ is the dimensionality, and $N_{\mathrm{r}}$ is the non-universal part of the occupation number. For the vacuum-to-superfluid transition in the two-dimensional Bose-Hubbard model, we have $N_{\mathrm{r}} = 0$ and $D=2$, and the predicted critical exponents are $z=2$ and $\nu=1/2$, characteristics of the dilute Bose gas universality class~\cite{Sachdev99QPT, Kaden10, Fisher89prb}.

\begin{figure}[t]
\includegraphics[width=2.9 in]{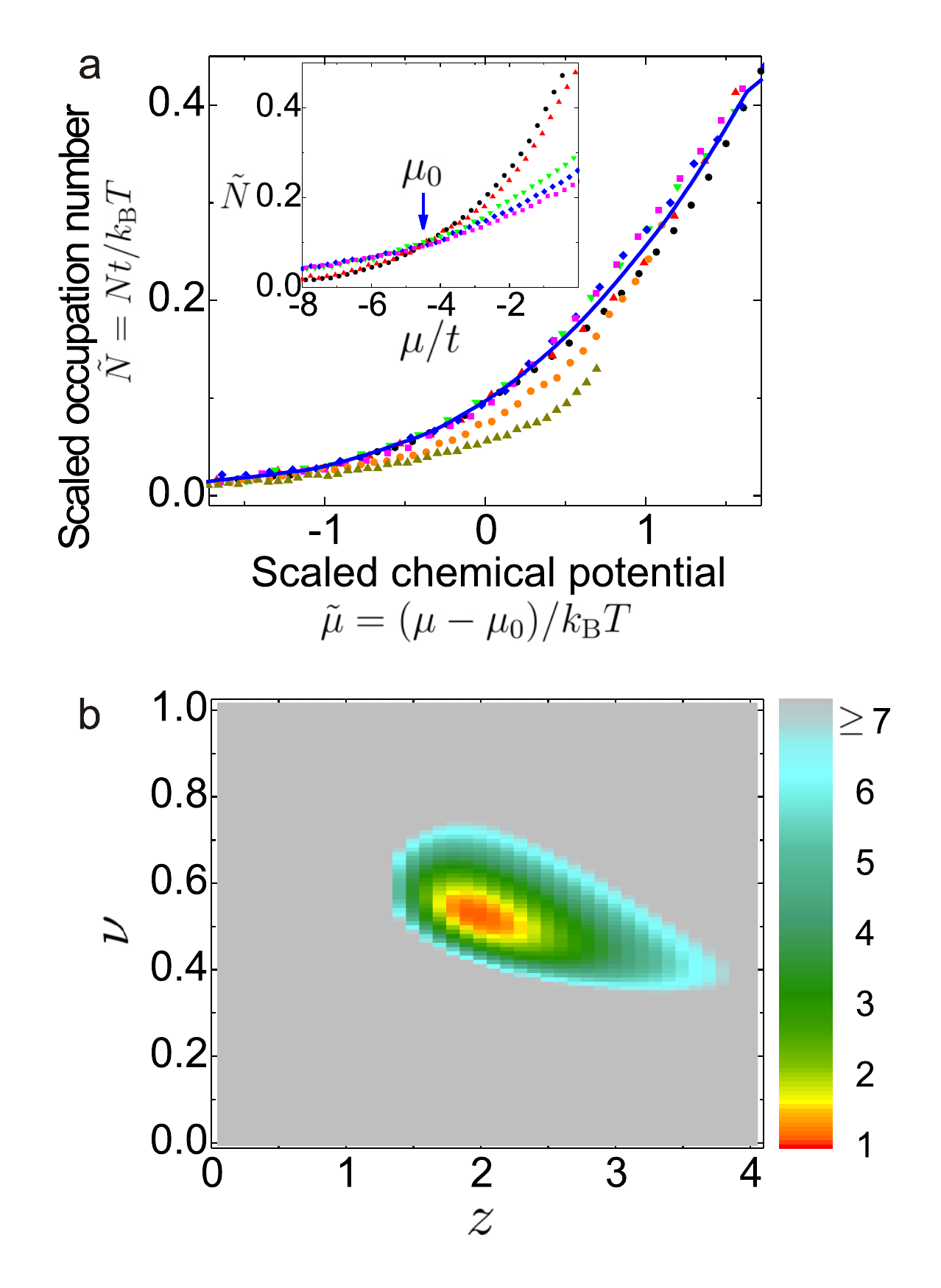}
\caption{\textbf{Evidence of a quantum critical regime.} \textbf{a,} Scaled occupation number $\tilde{N} = Nt/k_{\mathrm{B}}T$ as a function of the scaled chemical potential $\tilde{\mu} = (\mu-\mu_{0})/k_{\mathrm{B}}T$, measured at seven temperatures: $T = $5.8~nK (black circles), 6.7~nK (red triangles), 11~nK (green triangles), 13~nK (blue diamonds), 15~nK (magenta squares), 24~nK (orange circles), and 31~nK (dark yellow triangles), with the blue solid line showing the average curve for the lowest four temperatures. Inset shows the low-temperature data in the range of $T=5.8 \sim 15$~nK, and the critical chemical potential $\mu_{0}$ is identified as the crossing point; see text. The result, $\mu_{0} = -4.5(6) t$ agrees with the prediction $\mu_{0} = -4t$~\cite{Xibo11njp}. \textbf{b,} Determination of the dynamical critical exponent $z$ and the correlation length exponent $\nu$. The color represents the sum of normalized mean square deviations of $\tilde{N}$ for data in the low temperature range of $T=5.8 \sim 15$~nK, and indicates how well the scaled equation of state can collapse into one single curve (smaller deviations suggest better fits); see text. Best fit corresponds to $z = 2.0(3)$ and $\nu = 0.54(5)$. The uncertainties are determined from the contour on the $z$-$\nu$ plane where the deviation is $50\%$ above the minimum. The predicted values are $z=2$ and $\nu = 1/2$.} \label{fig2}
\end{figure}


Our experiment is based on 2D atomic gases of cesium-133 in 2D square optical lattices~\cite{Nate09Nat,CL10PRL}. The 2D trap geometry is provided by the weak horizontal~($r$-) confinement and strong vertical~($z$-) confinement, with envelope trap frequencies  $f_{\mathrm{r}} = 9.6$~Hz and $f_{\mathrm{z}} = 1940$~Hz, respectively. Details of the system and the sample preparation are described in Ref.~\cite{CL10PRL}. Typically $4,000$ to $20,000$ Bose-condensed atoms are loaded into the lattice. The lattice constant is $d = \lambda/2 = 0.532$ $\mu$m and the depth is $V_{\mathrm{L}} = 6.8$~$E_{\mathrm{R}}$, where $E_{\mathrm{R}} = k_{\mathrm{B}} \times 63.6$~nK is the recoil energy, $\lambda=1064$~nm is the lattice laser wavelength, and $h$ is the Planck constant. In the lattice, the tunneling energy is $t = k_{\mathrm{B}} \times 2.7$~nK, the on-site interaction is $U = k_{\mathrm{B}} \times 15$~nK, and the scattering length is $a = 15.9$~nm. The sample temperature is controlled in the range of $T$ = 5.8$\sim$31~nK.




We determine the equation of state $n(\mu,T)$ of the sample from the measured \textit{in situ} density distribution $n(x,y)$~\cite{CL11Nat, Nate09Nat}. The chemical potential $\mu(x,y)$ and the temperature $T$ are obtained by fitting the low-density tail of the sample where the atoms are normal. The fit is based on a mean-field model which accounts for interaction~\cite{Xibo11njp,Nikolay10njp}; see Methods.

We locate the quantum critical point by noting that at the critical chemical potential $\mu = \mu_{0}$, the scaled occupation number $\tilde{N} = Nt/k_{\mathrm{B}}T=nd^2t/k_{\mathrm{B}}T$ is temperature-independent, as indicated by Eq.~\ref{EoS}. We plot $\tilde{N}$ as a function of $\mu$ in the low temperature range of 5.8$\sim$15~nK, and indeed observe a crossing point at $\mu_{0} = -4.5(6)t$, see inset of Fig.~\ref{fig2}a. We identify this point as the critical point for the vacuum-to-superfluid transition, and our result agrees with the prediction $-4t$~\cite{Xibo11njp}.


\begin{figure}[t]
\includegraphics[width=3 in]{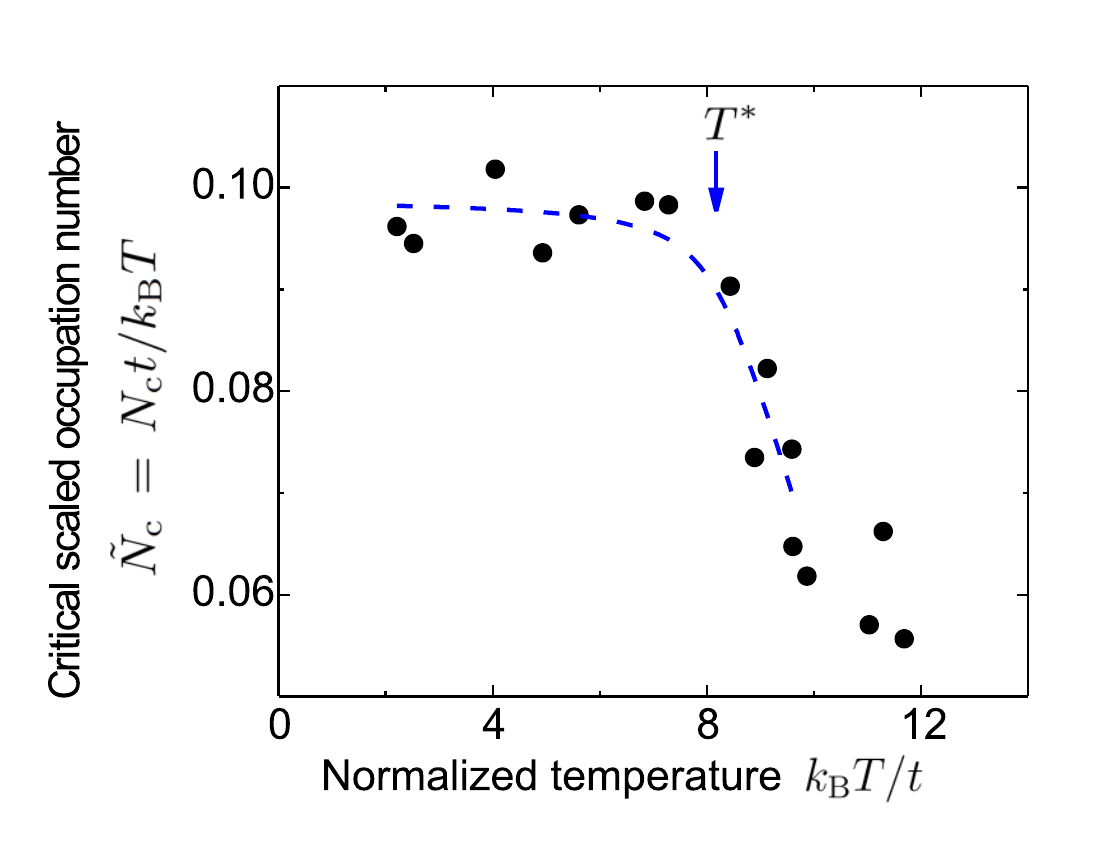}
\caption{\textbf{Finite-temperature effect on quantum critical scaling.} Scaled occupation number $\tilde{N}_{\mathrm{c}} = N_{\mathrm{c}}t/k_{\mathrm{B}}T$ at the critical chemical potential $\mu = \mu_{0}$ as a function of the normalized temperature $k_{\mathrm{B}}T/t$. The blue dashed line is an empirical fit, giving a temperature scale $T^{*} \approx 8t/k_{\mathrm{B}}$. For $T < T^{*}$, $\tilde{N}_{\mathrm{c}} \approx 0.097$ is independent of the temperature; for $T > T^{*}$, $\tilde{N}_{\mathrm{c}}$ deviates from the low-temperature value.  }\label{fig_Tstar}
\end{figure}

To test the critical scaling law, we compare the equation of state at different temperatures. Based on the expected exponents $z = 2$ and $\nu=1/2$, we plot the scaled occupation number $\tilde{N}$ as a function of the scaled chemical potential $\tilde{\mu} =
(\mu-\mu_{0})/k_{\mathrm{B}}T$; see Fig.~\ref{fig2}a. Below 15~nK, all the measurements collapse into a single curve, which confirms the emergence of the quantum critical scaling law (Eq.~\ref{EoS}) at low temperatures. Deviations become obvious at higher temperatures.

We determine the critical exponents $z$ and $\nu$ by comparing the equation of state at low temperatures. Taking various values of $z$ and $\nu$ in the range of $0<z<4$ and $0<\nu<1$, we compute the corresponding scaled occupation numbers $\tilde{N}$ and scaled chemical potentials $\tilde{\mu}$ based on Eq.~\ref{DefScaled}. We then evaluate how well the scaled equation of state in the range of $T=5.8 \sim 15$~nK can collapse to a single curve by computing the normalized mean-square deviations of $\tilde{N}$~(Methods). The result is shown in Fig.~\ref{fig2}b, from which we find the exponents to be $z=2.0(3)$ and $\nu=0.54(5)$. Our result is consistent with the predictions of the dilute Bose gas universality class: $z=2$ and $\nu=1/2$~\cite{Sachdev99QPT, Kaden10,Fisher89prb}, and shows the power of our method in identifying universality classes. In the following analyses, we adopt $z=2$ and $\nu = 1/2$.

\begin{figure}[t]
\includegraphics[width=2.6 in]{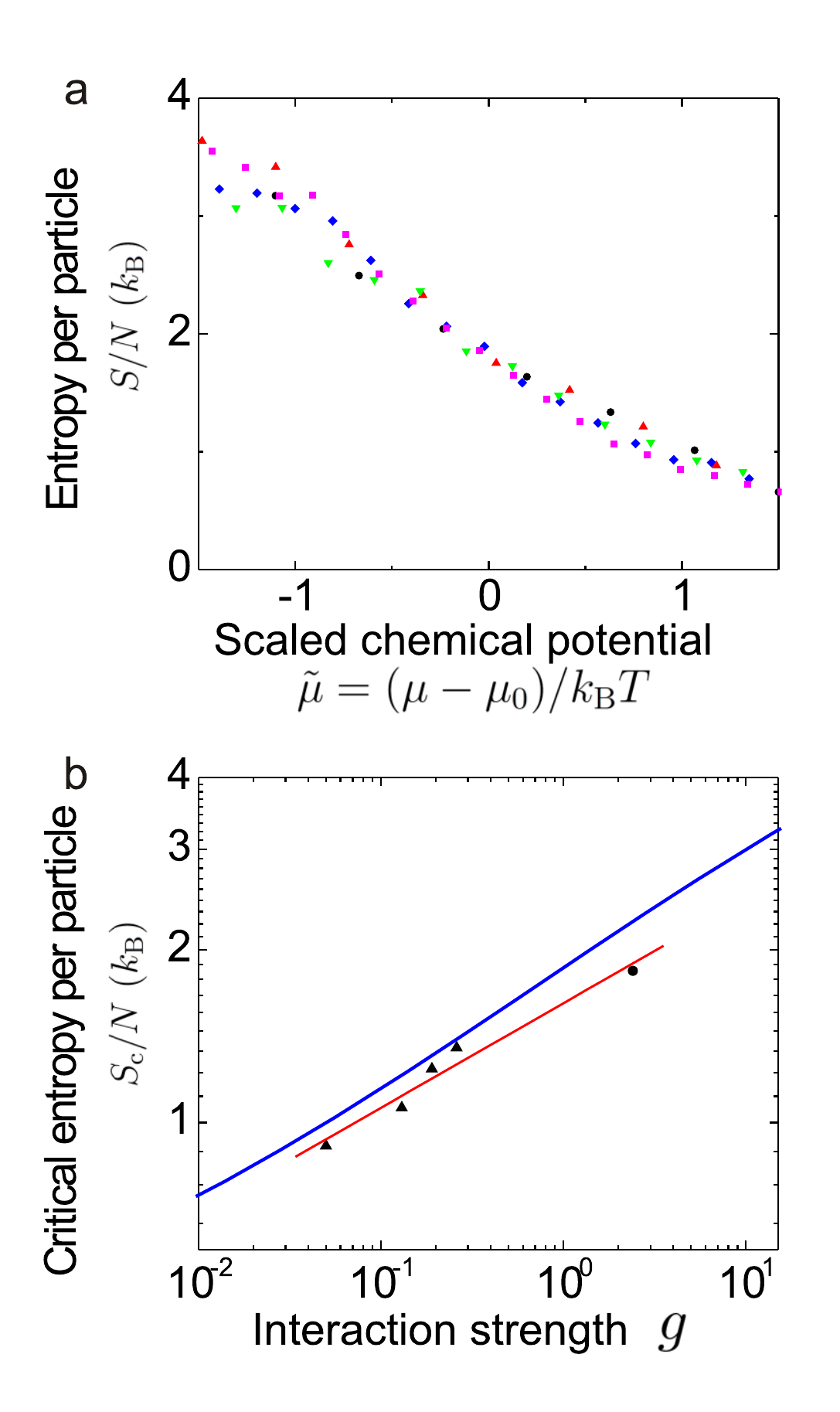}
\caption{\textbf{Entropy per particle in the critical regime.} 
\textbf{a,} Entropy per particle $S/N$ as a function of the scaled chemical potential $\tilde{\mu}$, measured in the temperature range of $5.8\sim 15$nK (same symbol and color scheme as in Fig.~\ref{fig2}a). \textbf{b,} Critical entropy per particle $S_{\mathrm{c}}/N$ as a function of the effective interaction strengths $g$: measurements for Bose gases with 2D optical lattices (black circle) and without lattice (black triangles, extracted from data in Ref.~\cite{CL11Nat}), mean-field calculations (blue line), and a power-law fit to the measurements, $S_{\mathrm{c}}/Nk_{\mathrm{B}} = 1.6(1) g^{0.18(2)}$(red line).}\label{fig_SovN}
\end{figure}

Our measurements at different temperatures allow us to investigate the breakdown of quantum criticality at high temperatures. To quantify the deviations, we focus on the temperature dependence of the scaled occupation number $\tilde{N}$ at the critical chemical potential $\mu = \mu_{0}$, as shown in Fig.~\ref{fig_Tstar}. Deviations from the low-temperature value are clear when the temperature exceeds $T^{*} = 22$~nK$\approx 8t/k_{\mathrm{B}}$. From this, we conclude that at $\mu = \mu_{0}$, the upper bound of thermal energy for the quantum critical behavior in our system is $k_{\mathrm{B}} T^{*} \approx 8t$. Our result is in fair agreement with the prediction of $6t$ based on quantum Monte Carlo calculations~\cite{DW10pra}.

From the equation of state, one can derive other thermodynamic quantities in the critical regime. Here we derive the entropy per particle $S/N$ based on measurements in the temperature range of $T=5.8\sim 15$~nK, using a procedure similar to that in Ref.~\cite{Dalibard11arxiv}. The measured entropy per particle only depends on the scaled chemical potential $\tilde{\mu}$ and monotonically decreases~(Fig.~\ref{fig_SovN}a), indicating a positive specific heat. Near the critical point $\tilde{\mu} = 0$, the entropy per particle has an approximate linear dependence on the scaled chemical potential: $S/Nk_{\mathrm{B}} = a-b\tilde{\mu}$, with $a = 1.8(1)$, $b=1.1(1)$. From this linear dependence, we derive an empirical thermodynamic relation analogous to the ideal gas law~(Methods):
\begin{equation}
P = C n^{x}\left(k_{\mathrm{B}}T\right)^{y},
\end{equation}
where $P$ is the pressure of the 2D gas, $x = \frac{2}{1+b} = 0.95(5)$, $y = \frac{2b}{1+b} = 1.05(5)$, $C = 0.8(2) (td^2)^{w} $ is a constant, and $w = \frac{1-b}{1+b} = -0.05(5)$.

Finally, we observe a weak dependence of the critical entropy per particle on the atomic interaction. Noting that a weakly-interacting 2D Bose gas follows similar scaling laws near $\mu = 0$~\cite{CL11Nat} due to the same underlying dilute Bose gas universality class~\cite{Sachdev99QPT,Sachdev06prb}, we apply similar analysis and extract the critical entropy per particle $S_{\mathrm{c}}/N$ at four interaction strengths $g \approx 0.05, 0.13, 0.19, 0.26$, as shown together with the lattice data ($g \approx 2.4$) in Fig.~\ref{fig_SovN}b. We observe a slow growing of $S_{\mathrm{c}}/N$ with $g$, and compare the measurements with mean-field calculations. The measured $S/N$ is systematically lower than the mean-field predictions, potentially due to quantum critical physics. The weak dependence on the interaction strength can be captured by a power-law fit to the data as $S_{\mathrm{c}}/Nk_{\mathrm{B}} = 1.6(1) g^{0.18(2)}$.

In summary, based on \textit{in situ} density measurements of Bose gases in 2D optical lattices, we confirm the quantum criticality near the vacuum-to-superfluid quantum phase transition. We show the suppression of the superfluid critical temperature, observe the universal scaling of equation of state, and extract the quantum critical point and critical exponents. In addition, we find that the entropy per particle is temperature-independent in the critical regime and has a weak dependence on the atomic interaction. Our experimental methods hold promise for identifying other quantum phase transitions, and prepare the tools for investigating quantum critical dynamics.  \\

\section*{Methods Summary}
Preparation and detection of cesium 2D Bose gases in optical lattices are similar to those described in Refs.~\cite{CL11Nat,CL10PRL}. We obtain the normal-to-superfluid transition point by collapsing the scaled equation of state at different temperatures~\cite{CL11Nat}. We extract the quantum critical points and critical exponents by minimizing the normalized mean square deviations of scaled occupation number. Based on the Gibbs-Duham equation, we derive thermodynamic quantities, in particular, the entropy per particle in the critical regime (details in Online Methods).

\noindent\\ \noindent \textbf{Acknowledgements}  We thank N. Prokof'ev and D.-W. Wang for discussions and numerical data, Q. Zhou, K. Hazzard, and N. Trivedi for discussions, and N. Gemelke for discussions and careful reading of the manuscript. The work was supported by NSF (grant numbers PHY-0747907, NSF-MRSEC DMR-0213745), the Packard foundation, and a grant from the Army Research Office with funding from the DARPA OLE program. 


\section*{Methods}
\noindent\textbf{Preparation and characterization of cesium 2D Bose gases in optical lattices} \emph{}

\noindent \emph{Preparation and detection of 2D gases in optical lattices}
The experimental procedure is similar to those described in Refs.~\cite{CL11Nat,CL10PRL}. We adjust the atomic temperature by applying magnetic field pulses near a magnetic Feshbach resonance~\cite{Chin10rmp} to excite the atoms~\cite{CL11Nat}. After the pulse, we tune the scattering length to $a=$ 15.9~nm, wait for 200 ms, and ramp on the optical lattice to 6.8~$E_{\mathrm{R}}$ in 270 ms. These parameters are chosen to allow the sample to reach thermal equilibrium after the ramp~\cite{CL10PRL}. The final lattice depth is sufficiently deep to validate single band Bose-Hubbard description. After preparing the sample, we perform \textit{in situ} absorption imaging using a strong resonant laser beam~\cite{CL11Nat,Dalibard11arxiv}. The atomic density is independently calibrated in similar methods as in Ref.~\cite{CL11Nat}.\\

\noindent \emph{Determination of the peak chemical potential and the temperature} 
We determine the peak chemical potential $\mu_{\mathrm{m}}$ (chemical potential at the trap center $r = 0$) and the temperature $T$ of a 2D Bose gas in 2D optical lattices by fitting the low-density tail of the azimuthally averaged density profile using the following formula~\cite{Xibo11njp}:
\begin{equation}\label{MFformula}
n(r) = d^{-2}\sum^{\infty}_{l=1}[I_0(2l\beta t)]^2 e^{l\beta[\mu_{\mathrm{m}} - 2U_{\mathrm{eff}}nd^2-V(r)]},
\end{equation}
This formula is based on local density approximation and a mean-field model that takes interaction into account. Here $n(r)$ is the 2D atomic density at radius $r$ from the cloud center, $d = 0.532$~$\mu$m is the lattice spacing, $I_0(x) = \int^{\pi}_{-\pi}\frac{\mathrm{d}\theta}{2\pi}e^{x\cos\theta}$ is the zeroth-order Bessel function with purely imaginary argument, $\beta
= 1/k_{\mathrm{B}}T$, $k_{\mathrm{B}}$ is the Boltzmann constant, $t$ is the tunneling, $V(r)$ is the envelope trapping potential, and $U_{\mathrm{eff}}$ is the effective interaction. Here the calculation of $U_{\mathrm{eff}}$ involves the Bose-Hubbard on-site interaction parameter $U$ and terms for a modified two-particle propagator~\cite{Nikolay10njp}:
\begin{equation}\label{Ueff}
U_{\mathrm{eff}} = \frac{U}{1+\frac{U}{2t}\Pi}, 
\end{equation}
where 
\begin{equation}
\Pi = \left(\frac{d}{2\pi}\right)^2\iint  \frac{\mathrm{d}k_{\mathrm{x}} \mathrm{d}k_{\mathrm{y}}}{\frac{k_{\mathrm{B}}T}{t} + 2\left[2-\cos(k_{\mathrm{x}}d) - \cos(k_{\mathrm{y}}d)\right]},\nonumber
\end{equation}
and the integration ranges of $k_{\mathrm{x}}$ and $k_{\mathrm{y}}$ are both from $-\pi/d$  to $\pi/d$, which covers the first Brillouin zone of the 2D square lattice. We test this formula on quantum Monte Carlo~(QMC) data~\cite{DW10pra}. Within our experimental temperature range, the fitted $T$ agrees with QMC value within $3\%$, and the fitted $\mu_{\mathrm{m}}$ agrees with QMC value within $0.6t$.\\

\noindent\textbf{Determination of the critical parameters} \emph{}

\noindent \emph{Normal-to-superfluid transition point $\mu_{\mathrm{c}}$} We use a procedure similar to that in Ref.~\cite{CL11Nat}. At a reference temperature $T_{\mathrm{ref}}$, we obtain the critical chemical potential $\mu_{\mathrm{c,ref}}$ for the normal-to-superfluid transition by fitting the crossover of the compressibility $\kappa = \frac{\mathrm{d}n}{\mathrm{d}\mu}$~(as a function of density $n$) near the transition region. The fit is based on an empirical formula $\kappa = kn-\sqrt{k^2(n-n_{\mathrm{c}})^2+w^2} +\sqrt{k^2n_{\mathrm{c}}^2 + w^2}$, where the slope $k$, critical density $n_{\mathrm{c}} =n(\mu_{\mathrm{ref}},T_{\mathrm{ref}})$, and the width $w$ of the transition region are fitting parameters. At a different temperature $T$, we obtain the transition point by comparing the equation of state at the two temperatures $T$ and $T_{\mathrm{ref}}$ and collapsing the scaled equation of state near the transition points into a single universal curve~\cite{CL11Nat}:  
\begin{equation}
\frac{n-n_{\mathrm{c}}}{k_{\mathrm{B}}T} = H(\frac{\mu - \mu_{\mathrm{c}}}{k_{\mathrm{B}}T}).
\end{equation}  
The resulting critical points are shown in Fig.~\ref{fig1}.\\

\noindent \emph{Quantum critical point $\mu_0$} For a given chemical potential $\mu$, we calculate the mean square deviation of the scaled occupation numbers $\tilde{N}=Nt/k_{\mathrm{B}}T$ at $M$ different temperatures, and normalize this deviation by the squared average value, as shown in the following formula:
\begin{equation}\label{deltaN_mu}
\Delta \tilde{N}(\mu) = \frac{1}{(M-1)\tilde{N}^2_{\mathrm{av}}(\mu)}\sum^M_{i=1}\left[\tilde{N}_i(\mu) - \tilde{N}_{\mathrm{av}}(\mu) \right]^2
\end{equation}
where $\tilde{N}_{\mathrm{av}}(\mu) = \sum^M_{i=1} \tilde{N}_i(\mu) / M$. The quantum critical point $\mu=\mu_0$ is determined by finding the minimum of $\Delta\tilde{N}(\mu)$.

\noindent \emph{Critical exponents $z$ and $\nu$} For given trial values of $z$ and $\nu$, we calculate the scaled occupation number $\tilde{N}$ and the scaled chemical potential $\tilde{\mu}$ according to Eq.~\ref{DefScaled}, and define the normalized mean square deviation, $\Delta\tilde{N}(\tilde{\mu})$, at a certain $\tilde{\mu}$ in a form similar to Eq.~\ref{deltaN_mu}. The exponents $z$ and $\nu$ are determined by minimizing the average $\Delta \tilde{N}(\tilde{\mu})$ in the range of $-1.5~k_{\mathrm{B}}T <\mu-\mu_{0}<1.5~k_{\mathrm{B}}T$.\\

\noindent\textbf{Universal thermodynamics in the quantum critical regime}
Based on the Gibbs-Duham equation~\cite{Qi10NatPhys}, we derive the pressure $P(\mu,T)$ from the \textit{in situ} density measurements:
\begin{eqnarray}   
P(\mu,T) &  = & \int^{\mu}_{-\infty} n(\mu',T) \mathrm{d}\mu'  
\end{eqnarray}
The entropy density $s(\mu,T)$ is related to the pressure via differentiation with respect to temperature:
\begin{eqnarray}
s(\mu,T) & = & \left(\frac{\partial P}{\partial T}\right)_{\mu} 
\end{eqnarray}

In the quantum critical regime near the vacuum-to-superfluid transition, the scale invariance of the equation of state suggests the following scaling laws for pressure $P$ and entropy density $s$:
\begin{eqnarray}   
P(\mu,T) 
 & =  & \left(\frac{k_{\mathrm{B}}T}{t}\right)^{\frac{D}{z}+1}K_P\left(\tilde{\mu}\right),
\end{eqnarray}  
\begin{eqnarray}\label{Eo_entropy}
s(\mu,T) 
 & = & \left(\frac{k_{\mathrm{B}}T}{t}\right)^{\frac{D}{z}}K_s\left(\tilde{\mu}\right),
\end{eqnarray}   
where $K_P$ and $K_s$ are generic functions of $\tilde{\mu}$. Combining Eq.~\ref{EoS} and Eq.~\ref{Eo_entropy}, we obtain the entropy per particle $S/N$ in unit of $k_{\mathrm{B}}$
\begin{eqnarray}
\frac{S}{Nk_{\mathrm{B}}} & = & 
\left(\frac{k_{\mathrm{B}}T}{t}\right)^{-1+\frac{1}{z\nu}}W\left(\tilde{\mu}\right)
\end{eqnarray}
where $W$ is a universal function.

For $z = 2$, $\nu=1/2$,                                                                                                              $S/Nk_{\mathrm{B}} = W(\tilde{\mu}) = 2K_P/{K_P}'-\tilde{\mu}$ is a temperature-independent function of the scaled chemical potential $\tilde{\mu} = (\mu-\mu_{0})/k_{\mathrm{B}}T$. In particular, near the critical point $\tilde{\mu} = 0$, $S/Nk_{\mathrm{B}}$ varies approximately as a linear function of $\tilde{\mu}$: $W(\tilde{\mu}) = a - b\tilde{\mu}$. Using $W(\tilde{\mu}) = 2K_P/{K_P}'-\tilde{\mu}$, we can solve the pressure $P$ and express it in terms of density $n$ and temperature $T$:
\begin{equation}
P = C n^x (k_{\mathrm{B}}T)^y,
\end{equation}
where $x = \frac{2}{1+b}$, $y = \frac{2b}{1+b}$, and the proportionality constant $C = (\frac{a}{2})^{\frac{2}{1+b}}(\frac{K_P(0)}{t^2})^{\frac{-1+b}{1+b}}$.

A similar technique was applied to obtain the entropy per particle for a bulk 2D gas of rubidium-87 atoms~\cite{Dalibard11arxiv}.\\

\noindent \emph{Effective interaction strength of a 2D gas}
We define the dimensionless effective interaction strength $g$ for our Bose gas in optical lattices: 
\begin{equation}\label{gDefinition}
g  =  U_{\mathrm{eff}}\frac{m^{*}d^2}{\hbar^2}
\end{equation}
and for that without lattices~\cite{CL11Nat}:
\begin{equation}
g = \frac{\sqrt{8\pi}a}{l_{z}}
\end{equation}
where the effective interaction $U_{\mathrm{eff}}$ is calculated using Eq.~\ref{Ueff}, $\hbar$ is the reduced Planck constant, $m^{*} = \hbar^2/E''(\mathrm{k})|_{\mathrm{k}=0}$ is the single-particle effective mass in a 2D optical lattice and can be calculated from the ground-band dispersion relation $E(\mathrm{k})$, $a$ is the scattering length which is tunable via a magnetic Feshbach resonance~\cite{Chin10rmp}, and $l_z$ is the vertical harmonic oscillator length.\\

\noindent \emph{Mean-field calculation on the entropy per particle}
We calculate the entropy per particle $S/N$ based on Eq.~\ref{MFformula}. At low temperatures $T$, the Bessel function takes its asymptotic form $I_0(x)\approx e^x/\sqrt{2 \pi x }$ when $x=2lt/(k_{\mathrm{B}}T)$ is large, and Eq.~\ref{MFformula} reduces to
\begin{equation}\label{MFzeroT}
F(\tilde{\mu}) = -\frac{1}{4\pi}\ln\left[1-\exp\left({\tilde{\mu}-\frac{2U_{\mathrm{eff}}}{t}F(\tilde{\mu})}\right)\right].\nonumber
\end{equation}
One can calculate $F(\tilde{\mu})$ by solving this equation self-consistently, and then derive $K_P(\tilde{\mu})$ and $S/Nk_{\mathrm{B}}= 2K_P/{K_P}'-\tilde{\mu}$ from $F(\tilde{\mu})$. In this calculation, the effective mass takes the value $m^{*}=\frac{\hbar^2}{2td^2}$ (under the tight-binding approximation); the effective interaction strength is thus given by $g = \frac{U_{\mathrm{eff}}}{2t}$ based on Eq.~\ref{gDefinition}.

\end{document}